\documentclass[11pt,english]{article}

\usepackage{hyperref,amsfonts,amsmath,amssymb,bm,a4wide,cite,subfigure,graphicx,babel}

\usepackage{mathabx}

\numberwithin{equation}{section}

\begin{document}

\title{\textbf{Quantum field theory treatment of the neutrino spin-flavor precession
in a magnetic field}}

\author{Maxim Dvornikov\thanks{maxim.dvornikov@gmail.com}
\\
\small{\ Pushkov Institute of Terrestrial Magnetism, Ionosphere} \\
\small{and Radiowave Propagation (IZMIRAN),} \\
\small{108840 Moscow, Troitsk, Russia}}

\date{}

\maketitle

\begin{abstract}
We study the spin-flavor precession of neutrinos in a magnetic field
within the quantum field theory approach in which neutrinos are virtual
particles. Neutrinos are taken to be Majorana particles having a nonzero
transition magnetic moment. We derive the dressed propagators of the
neutrino mass eigenstates exactly accounting for the magnetic field
contribution. The matrix element and the transition probability for the 
spin-flavor precession are obtained in the approximation of the forwardly scattered charged leptons. The leading term in the transition
probability is shown to coincide with the result of the standard quantum
mechanical description of neutrino oscillations. We also discuss the quantum field theory contributions to the neutrino dressed propagators and demonstrate that these contributions result in a small correction to the transition probability. The case of charged leptons with arbitrary energies is considered.
\end{abstract}

\section{Introduction}

The interaction of neutrinos with external fields can significantly
modify the dynamics of neutrino oscillations. From the point of view
of the phenomenology, the most important external field, contributing
to flavor oscillations, is the electroweak interaction of neutrinos
with background matter. It can result in the sizable amplification
of the transition probability, known as the Mikheyev--Smirnov--Wolfenstein (MSW)
effect~\cite{Wol78,MikSmi85}, that is the most plausible explanation
of the solar neutrino deficit~\cite{Xu23}.

Neutrinos are supposed to be electrically neutral particles. Nevertheless,
these particles can interact with electromagnetic fields owing to
nonzero magnetic moments which have a purely anomalous origin (see,
e.g., Ref.~\cite{LeeSch77}). It should be noted that the electromagnetic
properties of Dirac and Majorana neutrinos are completely different.
Dirac neutrinos can have both diagonal and transition magnetic moments,
whereas Majorana neutrinos can possess only transition ones. Despite
the issue of the neutrino nature is still open, neutrinos are likely
to be Majorana particles. At least, the smallness of the active neutrino
masses is explained more naturally for Majorana neutrinos~\cite{Kin04}.
The recent advances in the studies of neutrinos electromagnetic properties are
reviewed in Ref.~\cite{Giu24}.

If neutrinos with transition magnetic moments interact with an external
magnetic field, they undergo both flavor change and the spin flip.
This process is called the neutrino spin-flavor precession~\cite{LimMar88}.
The neutrino spin-flavor precession was studied, e.g., in Ref.~\cite{VolVysOku86}
to explain the possible anticorrelation between the solar activity
and the solar neutrinos flux. The spin-flavor precession accounting for the neutrino interaction with background matter was shown in Ref.~\cite{Akh88b} to exhibit a resonant behavior analogous to the MSW effect. The application of the predicted effect to solar neutrinos was also discussed in Ref.~\cite{Akh88b}.
Later, the spin-flavor precession was
understood in Ref.~\cite{AkhPul03} to play a subdominant role for
the solar neutrinos deficit problem. The description of the spin-flavor
precession of Majorana neutrinos expressed in terms of the Weyl spinors
was carried out in Ref.~\cite{SchVal81}.

As a rule, neutrino oscillations in vacuum and in external fields
are described within the quantum mechanical (QM) approach, which gives
satisfactory results in the majority of situations. The QM
formalism is based on the following assumptions. Flavor neutrinos are the coherent superposition of mass states. All mass states have definite equal momenta while evolving in time. The mass states are ultrarelativistic. The propagation time of a neutrino beam is approximately equal to the propagation distance. Nonetheless, these assumptions were shown in Ref.~\cite{NauNau20} to require the justification based upon a quantum field theory (QFT) formalism. Moreover, the effective wavefunction, which one deals with in QM approach, has to be somehow related to the true spinor wavefunctions of neutrinos. We should admit that the necessity of the QFT application to describe neutrino oscillations was realized long time ago. Numerous attempts to formulate a QFT based approach to neutrino oscillations are reviewed in Ref.~\cite{Beu03}.

We mention one of the QFT approaches which was put forward independently
in Refs.~\cite{Kob82,GiuKimLee93,GriSto96}. Neutrinos are supposed
to be virtual particles in this formalism. We also mention Refs.~\cite{CarChu99,AkhWil13},
where QFT was applied to study neutrino flavor oscillations in external
fields.

The approach in Refs.~\cite{Kob82,GiuKimLee93,GriSto96} allows one
to treat neutrino oscillations in vacuum since it involves the propagators
of the neutrino mass eigenstates. In vacuum, these propagators are
diagonal in the neutrino types. If a neutrino interacts with an external
field, the mass eigenstates propagators become nondiagonal since external fields mix different mass eigenstates as a rule. Finding
such nondiagonal propagators in not a trivial task. A particular case
of the spin-flavor precession of Dirac neutrinos, possessing diagonal
magnetic moments in the mass basis, was analyzed in frames of QFT
in Ref.~\cite{EgoVol22}.

Recently, in Ref.~\cite{Dvo25}, we developed a method which enables
one to find the propagators of neutrino mass eigenstates in an external
field. This approach is based on solving of the Dyson equations for
propagators accounting for the interaction with external fields. Using
this method in Ref.~\cite{Dvo25}, we described neutrino flavor oscillations
in background matter within QFT.

In the present work, we apply the results of Ref.~\cite{Dvo25} for
the studies of the neutrino spin-flavor precession in a magnetic field.
Moreover, we consider the situation when the interaction with an external
field is essentially nondiagonal in the mass basis. We shall see shortly
in Sec.~\ref{sec:ELECTRMAJ} that it is implemented for Majorana
neutrinos. Therefore, the application of the results of Ref.~\cite{Dvo25}
is crucial in this case. Note that the alternative formalism, based
on the relativistic quantum mechanics, for the description of neutrino
flavor oscillations and the spin-flavor precession in various external fields was
considered in Ref.~\cite{Dvo11}.

This work is organized in the following way. We remind briefly some
basic issues of the Majorana neutrinos electrodynamics in Sec.~\ref{sec:ELECTRMAJ}.
Then, in Sec.~\ref{sec:QFT}, we outline the QFT approach and adapt
it for the neutrino spin-flavor precession. We derive the dressed propagators
for the neutrino mass eigenstates in a magnetic field by solving the
corresponding Dyson equations in Sec.~\ref{sec:PROPB}. The matrix
element and the transition probability for the neutrino spin-flavor precession
are computed in Sec.~\ref{sec:TRANSPROB}. The results of the QFT
and the standard QM approaches are compared in Sec.~\ref{sec:SFOQM}.
Finally, we discuss our results in Sec.~\ref{sec:CONCL}. The propagators of
massive Weyl neutrinos in vacuum are derived in Appendix.

\section{Electrodynamics of Majorana neutrinos}\label{sec:ELECTRMAJ}

In this section, we remind how massive Majorana neutrinos interact
with electromagnetic fields.

Let us consider the system of two active flavor neutrinos $\nu_{\lambda}=(\nu_{e},\nu_{\mu})$.
The mass matrix of these neutrinos is supposed to be nondiagonal.
For simplicity, we take that no sterile neutrinos are present. To
diagonalize the mass matrix we introduce the mass eigenstates $\psi_{a}$,
$a=1,2$, by means of the matrix transformation,
\begin{equation}\label{eq:flmassrel}
  \nu_{\lambda}=\sum_{a}U_{\lambda a}\psi_{a},
  \quad
  (U_{\lambda a})=
  \left(
    \begin{array}{cc}
      \cos\theta & \sin\theta\\
      -\sin\theta & \cos\theta
    \end{array}
  \right),
\end{equation}
where $\theta$ is the vacuum mixing angle. We assume that $\psi_{a}$
are Majorana particles having masses $m_{a}$.

We suppose that $\psi_{a}$ have the nonzero matrix of magnetic moments
$(\mu_{ab})$. This matrix was found in Ref.~\cite{Pas00} to be
hermitian and purely imaginary. Hence, if we deal with two mass eigenstates,
we can represent this matrix as $\mu_{ab}=\mu(\sigma_{2})_{ab}$,
where $a,b=1,2$, $\sigma_{2}$ is the Pauli matrix, and $\mu$ is
the transition magnetic moment. Then, we suppose that these neutrinos
interact with an external electromagnetic field $F_{\mu\nu}=(\mathbf{E},\mathbf{B})$.
The wave equations for the mass eigenstates can be formally obtained
from the Dirac equation for $\psi_{a}$~\cite{Dvo12},
\begin{equation}\label{eq:Direq}
  (\mathrm{i}\gamma^{\mu}\partial_{\mu}-m_{a})\psi_{a}-\frac{\mu_{ab}}{2}\sigma_{\mu\nu}F^{\mu\nu}\psi_{b}=0,
\end{equation}
where $\gamma^{\mu}$, $\gamma^{5}$, and $\sigma_{\mu\nu}=(\mathrm{i}/2)(\gamma_{\mu}\gamma_{\nu}-\gamma_{\nu}\gamma_{\mu})$
are the Dirac matrices. Since the mass eigenstates are taken to be
Majorana particles, we represent them in terms of the Weyl spinors
$\eta_{a}$, $\psi_{a}^{\mathrm{T}}=(\mathrm{i}\sigma_{2}\eta_{a}^{*},\eta_{a})$.
Then, assuming that the electric field is absent and using Eq.~(\ref{eq:Direq}),
we derive the wave equations for the mass eigenstates,
\begin{align}\label{eq:Weyeqsys}
  \mathrm{i}\dot{\eta}_{1}+(\bm{\sigma}\mathbf{p})\eta_{1}+\mathrm{i}m_{1}\sigma_{2}\eta_{1}^{*}
  -\mu(\bm{\sigma}\mathbf{B})\sigma_{2}\eta_{2}^{*} & =0,
  \nonumber
  \\
  \mathrm{i}\dot{\eta}_{2}+(\bm{\sigma}\mathbf{p})\eta_{2}+\mathrm{i}m_{2}\sigma_{2}\eta_{2}^{*}
  +\mu(\bm{\sigma}\mathbf{B})\sigma_{2}\eta_{1}^{*} & =0,
\end{align}
where $\mathbf{p}=-\mathrm{i}\nabla$ is the momentum operator and
$\bm{\sigma}=(\sigma_{1},\sigma_{2},\sigma_{3})$ are the Pauli matrices.

One can see in Eq.~(\ref{eq:Weyeqsys}) that the neutrino magnetic
interaction causes transitions between different mass eigenstates,
$1\leftrightarrow2$. Moreover, the magnetic interaction has the structure
similar to the mass terms, which are known to induce neutrino-to-antineutrino
oscillations~\cite{Kob82,SchVal80}. Indeed, both terms are $\propto\sigma_{2}\eta_{a,b}^{*}$.
Thus, neutrinos undergo transitions $1\leftrightarrow\bar{2}$ and
$\bar{1}\leftrightarrow2$, where the bar denotes an antiparticle
state, in an external magnetic field.

\section{Quantum field theory formalism for neutrino oscillations}\label{sec:QFT}

In this section, we briefly outline the quantum field theory based
formalism for neutrino oscillations in which neutrinos are virtual
particles~\cite{Kob82,GiuKimLee93,GriSto96}. We adapt this method
for the description of the spin-flavor precession.

\begin{figure}
  \centering
  \includegraphics{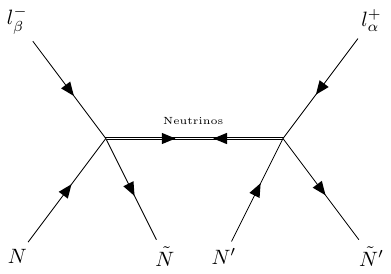}
  \protect
  \caption{The schematic Feynman diagram for the reaction $l_{\beta}^{-}+N\to\text{neutrinos}\to l_{\alpha}^{+}+\tilde{N}'$.
  The arrows over fermionic lines, $l_\beta^-$, $l_\alpha^+$, and neutrinos, show the flow of the lepton number.
  Since we do not care about the details of the nuclear transformations in the source and the detector,
  the arrows over nuclear lines, $N$, $\tilde{N}$, $N'$, and $\tilde{N}'$, can signify the baryon number flow.
  The neutrino line has two counter-directional arrows since we deal
  with the Majorana particles which violate the lepton number. The time flows from left to right.
  Moreover, the neutrino line is broad
  since the mass eigenstates interact with a magnetic field [see Eq.~(\ref{eq:Weyeqsys})].
  Note that we take into account the magnetic interaction exactly (see Sec.~\ref{sec:PROPB}).\label{fig:schemsfo}}
\end{figure}

Following Ref.~\cite{Kob82}, we assume that the source and the detector
of neutrinos, separated by the distance $\mathbf{L}$, consist of
heavy nuclei $N$ and $N'$ at $t\to -\infty$. In addition to these nuclei, we have the charged lepton $l_{\beta}^{-}$, where $\beta=e,\mu,\dotsc$, in the in-state. Since the charge lepton has a definite mass, its in-state is well defined. Then, we take that $l_{\beta}^{-}$ interacts with the source, i.e. with the nucleus $N$. Neutrino states are produced in this interaction: $l_{\beta}^{-} + N \to \tilde{N} + \text{neutrinos}$. Note that a nuclear transformation takes place in a source, i.e. $N \neq \tilde{N}$, since electric charges of $l_{\beta}^{-}$ and neutrinos are different. The interaction of $l_{\beta}^{-}$ with the source corresponds to the left hand vertex in Fig.~\ref{fig:schemsfo}.

Then, the neutrino beam travels towards a detector consisting of a nucleus $N'$. While reaching it, neutrinos are absorbed by $N'$: $\text{neutrinos} + N' \to l_{\alpha}^{+} + \tilde{N}'$, where $l_{\alpha}^{+}$ is the anti-lepton of the flavor $\alpha=e,\mu,\dotsc$ and $\tilde{N}'$ is the new nucleus different from $N'$. Therefore, at $t\to + \infty$ one has $\tilde{N}$ in a source, $\tilde{N}'$ in a detector, as well as $l_{\alpha}^{+}$. Analogously to $l_{\beta}^{-}$, the anti-lepton $l_{\alpha}^{+}$ also has a definite mass. Thus, the out-state of $l_{\alpha}^{+}$ at $t\to+\infty$ is also well defined. The neutrino interaction with a source corresponds to the vertex in the right hand side in Fig.~\ref{fig:schemsfo}.

Now, we suppose that the flavor content of charged leptons changes, i.e. $\beta\neq\alpha$. In this situation, one can speak about neutrino flavor oscillations. Moreover, we assume that electric charges of leptons are opposite. For example, we have an electron in a source and
antimuon in a detector. Such a process is allowed in vacuum only for Majorana
neutrinos.  It was connected to neutrino-to-antineutrino oscillations
in Refs.~\cite{Kob82,SchVal80}. That is why we place counter-directional arrows over the neutrino line in Fig.~\ref{fig:schemsfo} to emphasize that the lepton number is violated. However, if we study the transition
$l_{\beta}^{-}+N\to\text{neutrinos}\to l_{\alpha}^{+}+\tilde{N}'$ in an external
field which neutrinos interact with, it is more appropriate to call
it the neutrino spin-flavor precession.

The $S$-matrix element of the reaction in question reads~\cite{Kob82}
\begin{equation}\label{eq:Smatrgen}
  S=-\frac{1}{2}
  \left(
    \sqrt{2}G_{\mathrm{int}}
  \right)^{2}
  \int\mathrm{d}^{4}x\mathrm{d}^{4}y
  \Big\langle
    l_{\alpha}^{+},\tilde{N},\tilde{N}'
  \Big|
  T
  \left\{
    j^{\mu}(x)J_{\mu}(x)j^{\nu}(y)J_{\nu}(y)
  \right\}
  \Big|
    l_{\beta}^{-},N,N'
  \Big\rangle,
\end{equation}
where $G_{\mathrm{int}}$ is the coupling constant,
\begin{equation}\label{eq:lepcurr}
  j_{\mu}=\sum_{\lambda}\bar{\nu}_{\lambda}\gamma_{\mu}^{\mathrm{L}}l_{\lambda},
\end{equation}
is the leptons current, $\gamma_{\mu}^{\mathrm{L}}=\gamma_{\mu}(1-\gamma^{5})/2$,
and $J_{\mu}(x)$ is the nuclear current operator. Then, we represent
the operator valued wave function of leptons as
\begin{equation}\label{eq:lepplanewaves}
  l_{\lambda}(x)=\int\frac{\mathrm{d}^{3}p}{(2\pi)^{3/2}}
  \left[
    a_{\lambda}(p)u_{\lambda}(p)e^{-\mathrm{i}p_{\lambda}x}+b_{\lambda}^{\dagger}(p)v_{\lambda}(p)e^{\mathrm{i}p_{\lambda}x}
  \right],
\end{equation}
where $b_{\lambda}^{\dagger}(p)$ and $a_{\lambda}(p)$ are the creation
and annihilation operators of antileptons and leptons, $u_{\lambda}(p)$
and $v_{\lambda}(p)$ are the basis spinors, $p_{\lambda}^{\mu}=(E_{\lambda},\mathbf{p})$,
$E_{\lambda}=\sqrt{\mathfrak{m}_{\lambda}^{2}+p^{2}}$ is the lepton
energy, and $\mathfrak{m}_{\lambda}$ is its mass. We take that $\left|l_{\beta}^{-}\right\rangle =a_{\beta}^{\dagger}(p_{\beta})\left|0\right\rangle $
and $\left\langle l_{\alpha}^{+}\right|=\left\langle 0\right|b_{\alpha}(p_{\alpha})$,
i.e. we deal with the plane waves incoming and outgoing leptons. Eventually,
we rewrite Eq.~(\ref{eq:Smatrgen}) in the following way:
\begin{align}\label{eq:Smatrint}
  S= & 2G_{\mathrm{int}}^{2}
  \sum_{ab}U_{\alpha a}^{*}U_{\beta b}^{*}v_{\alpha}^{\mathrm{T}}(p_{\alpha})(\gamma_{\mathrm{L}}^{\mu})^{\mathrm{T}}
  \nonumber
  \\
  & \times
  \left(
    \int\mathrm{d}^{4}x\mathrm{d}^{4}ye^{\mathrm{i}p_{\alpha}x-\mathrm{i}p_{\beta}y}J_{\mu}(x)J_{\nu}(y)
    \big\langle
      0
    \big|
    T
    \left\{
      (\bar{\psi}_{a}(x))^{\mathrm{T}}\bar{\psi}_{b}(y)
    \right\}
    \big|
      0
    \big\rangle
  \right)
  \gamma_{\mathrm{L}}^{\nu}u_{\beta}(p_{\beta}),
\end{align}
where we account for Eqs.~(\ref{eq:lepcurr}) and~(\ref{eq:lepplanewaves}). In Eq.~\eqref{eq:Smatrint}, we use the mean values of the nuclear currents $J_{\mu}(x) = \langle \tilde{N}' | J_\mu | N' \rangle$ and $J_{\nu}(y) = \langle \tilde{N} | J_\nu | N \rangle$.

We choose the simplest model of the source and the detector.  We assume that the nuclei $N$ and $N'$, as well as $\tilde{N}$ and $\tilde{N}'$, are quite
heavy. Hence, their positions, $\mathbf{x}_{1}$ and $\mathbf{x}_{2}$,
are unchanged, with $\mathbf{x}_{2}-\mathbf{x}_{1}=\mathbf{L}$. In
this case, the nuclear currents in Eq.~(\ref{eq:Smatrint}) have
the form,
\begin{equation}\label{eq:nuclmatrel}
  J_{\mu}(x_{0},\mathbf{x})\propto\delta_{\mu0}\delta(\mathbf{x}-\mathbf{x}_{2}),
  \quad
  J_{\nu}(y_{0},\mathbf{y})\propto\delta_{\nu0}\delta(\mathbf{y}-\mathbf{x}_{1}).
\end{equation}
i.e. the nuclear matrix elements are time independent. Therefore, the consideration of the source and the detector on the classical level allows one to factorize the nuclear contribution in the $S$-matrix in Eq.~\eqref{eq:Smatrint}.

We take that the neutrino mass eigenstates are Majorana particles,
i.e. $\left(\psi_{a}\right)^{c}=C\bar{\psi}_{a}^{\mathrm{T}}=\psi_{a}$,
where $C$ is the charge conjugation matrix. In this case,
\begin{multline}\label{eq:proprel}
  v_{\alpha}^{\mathrm{T}}(p_{\alpha})(\gamma_{\mathrm{L}}^{0})^{\mathrm{T}}
  \left\langle 0\left|T\left\{ (\bar{\psi}_{a}(x))^{\mathrm{T}}\bar{\psi}_{b}(y)\right\} \right|0\right\rangle
  \gamma_{\mathrm{L}}^{0}u_{\beta}(p_{\beta})
  \\
  =\kappa_{\bar{\alpha}\mathrm{R}}^{\dagger}(p_{\alpha})
  \left\langle 0\left|T\left\{ \xi_{a}(x)\eta_{b}^{\dagger}(y)\right\} \right|0\right\rangle
  \kappa_{\beta\mathrm{L}}(p_{\beta}).
\end{multline}
Here we use the representations
\begin{equation}\label{eq:spinrepr}
  \psi_{a}=
  \left(
    \begin{array}{c}
      \xi_{a}\\
      \eta_{a}
    \end{array}
  \right),
  \quad
  u_{\alpha,\beta}(p_{\alpha,\beta})=
  \left(
    \begin{array}{c}
      \kappa_{\bar{\alpha},\beta\mathrm{R}}(p_{\alpha,\beta})\\
      \kappa_{\bar{\alpha},\beta\mathrm{L}}(p_{\alpha,\beta})
    \end{array}
  \right).
\end{equation}
Note that we keep $\xi_{a}$ instead of $\mathrm{i}\sigma_{2}\eta_{a}^{*}$
in Eq.~(\ref{eq:spinrepr}). We also use the known expressions, $\gamma_{\mathrm{R}}^{\mu}=C(\gamma_{\mathrm{L}}^{\mu})^{\mathrm{T}}C$
and $v_{\alpha}(p_{\alpha})=\left(u_{\alpha}(p_{\alpha})\right)^{c}$,
to derive Eq.~(\ref{eq:proprel}).

Using Eq.~(\ref{eq:proprel}) and performing the integrations as
in Ref.~\cite{Kob82}, we represent Eq.~(\ref{eq:Smatrint}) in
the form,
\begin{equation}\label{eq:Smatrencons}
  S=4\pi G_{\mathrm{int}}^{2}e^{-i\mathbf{p}_{\alpha}\mathbf{x}_{2}+i\mathbf{p}_{\beta}\mathbf{x}_{1}}\delta(E_{\alpha}-E_{\beta})
  \mathcal{M}_{\beta\to\bar{\alpha}}.
\end{equation}
One can see in Eq.~\eqref{eq:Smatrencons} that the total energy is conserved, $E_\alpha = E_\beta$. This fact results from the assumption that nuclei in the source and the detector are static. Moreover, formally, we consider the situation like a scattering, i.e. all external particles are in- and out-states at $t \to \mp \infty$. Hence, one has the infinite observation time $-\infty < t < +\infty$. It allows us to integrate over the infinite time intervals in Eq.~\eqref{eq:Smatrint} to get the energy conservation delta-function in Eq.~\eqref{eq:Smatrencons}. The violation of the energy conservation assumption would lead to the decoherence in neutrino oscillation, which is discussed shortly in Sec.~\ref{sec:CONCL}.

The matrix element in Eq.~\eqref{eq:Smatrencons} is
\begin{equation}\label{eq:matrelgen}
  \mathcal{M}_{\beta\to\bar{\alpha}}=\sum_{ab}U_{\alpha a}^{*}U_{\beta b}^{*}\kappa_{\bar{\alpha}\mathrm{R}}^{\dagger}(p_{\alpha})
  \left(
    \int\frac{\mathrm{d}^{3}q}{(2\pi)^{3}}\Sigma_{ab}(E,\mathbf{q})e^{\mathrm{i}\mathbf{qL}}
  \right)
  \kappa_{\beta\mathrm{L}}(p_{\beta}).
\end{equation}
Here, $E=(E_{\alpha}+E_{\beta})/2$ and $\Sigma_{ab}(q_{0},\mathbf{q})$
are the 4D Fourier images of the nondiagonal propagators,
\begin{equation}\label{eq:propx}
  \Sigma_{ab}(x-y)=
  \left\langle
    0
  \left|
  T
  \left\{
    \xi_{a}(x)\eta_{b}^{\dagger}(y)
  \right\}
  \right|
    0
  \right\rangle.
\end{equation}
If we consider utrarelativistic incoming and outgoing leptons, then,
in Eq.~(\ref{eq:matrelgen}), we should replace $\kappa_{\beta\mathrm{L}}$
and $\kappa_{\bar{\alpha}\mathrm{R}}$ with the helicity states $\kappa_{\beta-}$
and $\kappa_{\bar{\alpha}+}$.

The probability of the neutrino spin-flavor precession is $P_{\beta\to\bar{\alpha}}\propto|\mathcal{M}_{\beta\to\bar{\alpha}}|^{2}$.
In case of the Majorana neutrinos motion in vacuum, the propagators
are diagonal, $\Sigma_{ab}\propto\delta_{ab}$. In this situation,
the process $\beta\to\bar{\alpha}$ corresponds to neutrino-to-antineutrino
oscillations, as we mentioned earlier. The probability of $\nu_{\alpha}\to\bar{\nu}_{\beta}$
oscillations was found in Refs.~\cite{Kob82,SchVal80} to be of the
order $\sim(m_{1,2}/E)^{2}$.

If Majorana neutrinos have a transition magnetic moment, as one can
see in Eq.~(\ref{eq:Weyeqsys}), the wave equations for $\eta_{1}$
and $\eta_{2}$ are coupled by the magnetic interaction. This interaction
induces the transitions between $\eta_{1,2}$ and $\mathrm{i}\sigma_{2}\eta_{2,1}^{*}=\xi_{2,1}$.
It means that both diagonal, $\Sigma_{11}$ and $\Sigma_{22}$, and
nondiagonal, $\Sigma_{12}$ and $\Sigma_{21}$, propagators in Eq.~(\ref{eq:propx})
can contribute to the transition probability. In case we study the spin-flavor precession like $\nu_{e}\to\bar{\nu}_{\mu}$, the matrix element
in Eq.~(\ref{eq:matrelgen}) takes the form,
\begin{equation}\label{eq:matrelemu}
  \mathcal{M}_{e\to\bar{\mu}}=\kappa_{\bar{\mu}\mathrm{R}}^{\dagger}(p_{\mu})\int\frac{\mathrm{d}^{3}q}{(2\pi)^{3}}e^{\mathrm{i}\mathbf{qL}}
  \left[
    \frac{1}{2}\left(\Sigma_{22}-\Sigma_{11}\right)\sin2\theta+\cos^{2}\theta\Sigma_{21}-\sin^{2}\theta\Sigma_{12}
  \right]
  \kappa_{e\mathrm{L}}(p_{e}),
\end{equation}
where we use Eq.\ (\ref{eq:flmassrel}). Our main goal is to find
the propagators $\Sigma_{ab}$ in Eq.~(\ref{eq:matrelemu}) in the
presence of the magnetic field.

Earlier, we mentioned that the marker of the spin-precession is the appearance of the anti-lepton $l_\alpha^+$ (out-state) in the detector provided one had the lepton $l_\beta^-$ (in-state) in the source. It leads to the bi-spinors $v_\alpha$ and $u_\beta$, as well as the propagator of massive neutrinos $\propto \left\langle 0\left|T\left\{ (\bar{\psi}_{a}(x))^{\mathrm{T}}\bar{\psi}_{b}(y)\right\} \right|0\right\rangle$ in Eq.~\eqref{eq:Smatrint}. Note that this propagator is nonzero only when neutrinos are Majorana particles. That is why, virtual neutrinos undergo the transitions $\nu_{\beta} \to \bar{\nu}_{\alpha}$ effectively. Since only left neutrinos are active, we may say that we deal with the transitions $\nu_{\beta\mathrm{L}} \to \bar{\nu}_{\alpha\mathrm{L}}$.

If we studied the analogue of the process in Fig.~\ref{fig:schemsfo} for Dirac neutrinos, one would have leptons $l_{\alpha,\beta}^-$ with the same electric charge interacting with a detector and a source (out- and in-states). Therefore, in Eq.~\eqref{eq:Smatrint}, we would have the bi-spinors $u_{\alpha,\beta}$, as well as the propagator $\propto \left\langle 0\left|T\left\{ \psi_{a}(x)\bar{\psi}_{b}(y)\right\} \right|0\right\rangle$ for Dirac neutrinos. This situation would correspond to the effective transformation of virtual neutrinos like $\nu_{\beta} \to \nu_{\alpha}$ or $\nu_{\beta\mathrm{L}} \to \nu_{\alpha\mathrm{L}}$.

Finally, we mention that despite the similarity in the description of spin-flavor precession of Majorana and Dirac neutrinos (see, e.g., Refs.~\cite{LimMar88,Akh88b}), there is a conceptual difference between these two processes. The spin-flavor precession of Dirac particles allows a quasiclassical description since one can introduce an effective spin vector which precesses in an external magnetic field~\cite{EgoLobStu00}. For Majorana neutrinos, it is the essentially quantum process since it involves transitions between particles and antiparticles.

\section{Propagators of massive Majorana neutrinos in a magnetic field}\label{sec:PROPB}

In this section, we derive the dresses propagators
of Majorana neutrinos interacting with a magnetic field. For this
purpose, we use the method developed in Ref.~\cite{Dvo25}, where
the propagators in question are the solutions of the Dyson equations
exactly accounting for the interaction with an external field. First,
we derive the Dyson equations within the perturbative approach, with
all the terms in the perturbation series being summed up. Then, the
Dyson equations are solved exactly for ultrarelativistic neutrinos.
We deal with two mass eigenstates since we mentioned in Ref.~\cite{Dvo25}
that extension of the formalism for the greater number of neutrinos
is ambiguous.

\begin{figure}[htbp]
  \centering
  \subfigure[]
  {\label{fig:f1a}
  \includegraphics[viewport=150bp 600bp 450bp 730bp,clip,scale=0.65]{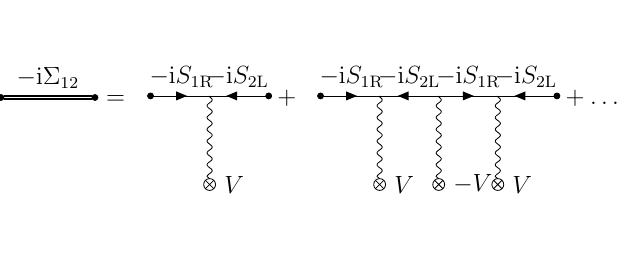}}
  \subfigure[]
  {\label{fig:f1b}
  \includegraphics[viewport=150bp 600bp 450bp 730bp,clip,scale=0.65]{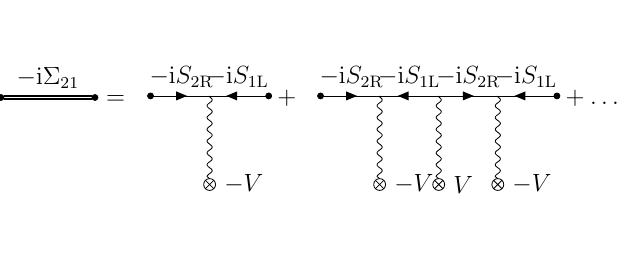}}
  \\
  \subfigure[]
  {\label{fig:f1c}
  \includegraphics[viewport=150bp 600bp 450bp 730bp,clip,scale=0.65]{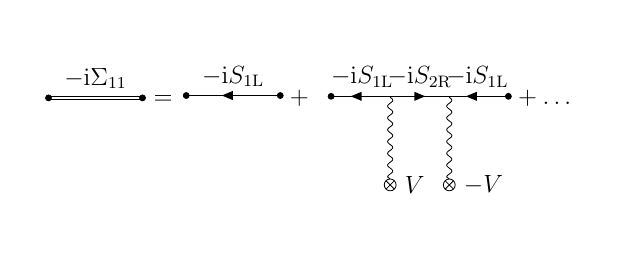}}
  \subfigure[]
  {\label{fig:f1d}
  \includegraphics[viewport=150bp 600bp 450bp 730bp,clip,scale=0.65]{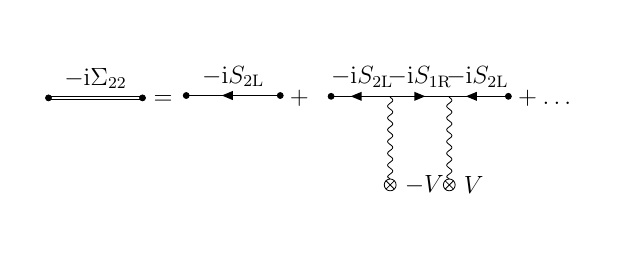}}
  \protect 
\caption{Feynman diagrams corresponding to Eq.~(\ref{eq:Dysser}).
4D Fourier images of the bare propagators $-\mathrm{i}S_{a\mathrm{R,L}}$, given in Eqs.~\eqref{eq:SL} and~\eqref{eq:SR}, are
depicted by thin solid lines~\cite[pp.~468--472]{BerLifPit82}. The broad solid lines correspond
to the dressed propagators $-\mathrm{i}\Sigma_{ab}$. The potential of the magnetic interaction is $V=\mathrm{i}\mu(\bm{\sigma}\mathbf{B})$.\label{fig:Dyson}}
\end{figure}

The magnetic interaction in Eq.~(\ref{eq:Weyeqsys}) mixes both opposite
helicities and different mass eigenstates. Thus, if we have a left-handed
lepton in a source and a right-handed one in a detector, like in Eq.~(\ref{eq:matrelemu}),
the dressed propagators are the sums of the following perturbative
series:
\begin{align}\label{eq:Dysser}
  -\mathrm{i}\Sigma_{12}= & (-\mathrm{i}S_{1\mathrm{R}})V(-\mathrm{i}S_{2\mathrm{L}})
  +(-\mathrm{i}S_{1\mathrm{R}})V(-\mathrm{i}S_{2\mathrm{L}})(-V)(-\mathrm{i}S_{1\mathrm{R}})V(-\mathrm{i}S_{2\mathrm{L}})+\dotsb,
  \nonumber
  \\
  -\mathrm{i}\Sigma_{21}= & (-\mathrm{i}S_{2\mathrm{R}})(-V)(-\mathrm{i}S_{1\mathrm{L}})
  +(-\mathrm{i}S_{2\mathrm{R}})(-V)(-\mathrm{i}S_{1\mathrm{L}})V(-\mathrm{i}S_{2\mathrm{R}})(-V)(-\mathrm{i}S_{1\mathrm{L}})+\dotsb,
  \nonumber
  \\
  -\mathrm{i}\Sigma_{11}= & -\mathrm{i}S_{1\mathrm{L}}
  +(-\mathrm{i}S_{1\mathrm{L}})V(-\mathrm{i}S_{2\mathrm{R}})(-V)(-\mathrm{i}S_{1\mathrm{L}})+\dotsb,
  \nonumber
  \\
  -\mathrm{i}\Sigma_{22}= & -\mathrm{i}S_{2\mathrm{L}}
  +(-\mathrm{i}S_{2\mathrm{L}})(-V)(-\mathrm{i}S_{1\mathrm{R}})V(-\mathrm{i}S_{2\mathrm{L}})+\dotsb,
\end{align}
where $V=\mathrm{i}\mu(\bm{\sigma}\mathbf{B})$ is the magnetic interaction
and $S_{a\mathrm{L,R}}$ are the propagators of left-handed neutrinos
and right-handed antineutrinos. Since the magnetic interaction contributes
to the transitions $a\leftrightarrow\bar{b}$ only, the diagonal propagators
$S_{a\mathrm{L,R}}$ should be taken in vacuum. We give them in Eqs.~(\ref{eq:SL})
and~(\ref{eq:SR}). We also take into account in Eq.~(\ref{eq:Dysser})
that the magnetic interaction contributes to the wave equations for
$\eta_{1}$ and $\eta_{2}$ with opposite signs; cf. Eq.~(\ref{eq:Weyeqsys}).
The perturbative series in Eq.~(\ref{eq:Dysser}) are illustrated
in Fig.~\ref{fig:Dyson}.

The series in Eq.~(\ref{eq:Dysser}) can be summed up, giving the result
\begin{align}\label{eq:Dyseq}
  \mathrm{i}\Sigma_{12}^{-1} & =-(S_{1\mathrm{R}}VS_{2\mathrm{L}})^{-1}+V,
  \quad
  \mathrm{i}\Sigma_{21}^{-1} =(S_{2\mathrm{R}}VS_{1\mathrm{L}})^{-1}-V,
  \nonumber
  \\
  \Sigma_{11}^{-1} & =S_{1\mathrm{L}}^{-1}-VS_{2\mathrm{R}}V,
  \quad
  \Sigma_{22}^{-1} =S_{2\mathrm{L}}^{-1}-VS_{1\mathrm{R}}V.
\end{align}
We notice the diagonal propagators in Eq.~(\ref{eq:Dyseq}) do not
contribute to the matrix element since $\Sigma_{aa}$ preserves the neutrino helicity, whereas $\mathcal{M}_{e\to\bar{\mu}}$ implies the spin-flip.
We provide the detailed calculation for $\Sigma_{12}$
and $\Sigma_{21}$ only.

One can see in the expressions for $S_{a\mathrm{L,R}}$ in Eqs.~(\ref{eq:SL})
and~(\ref{eq:SR}) that these propagators contain the projectors
$(1\mp\bm{\sigma}\hat{p})$ for ultrarelativistic neutrinos. Thus,
the inverse operators $S_{a\mathrm{L,R}}^{-1}$, which are present
in Eq.~(\ref{eq:Dyseq}), do not exist for such neutrinos. However,
we keep the ratio $\varsigma_{a}=p/E_{a}$ in soling of Eq.~(\ref{eq:Dyseq})
and set $\varsigma_{a}\to1$ in the final result.


We provide the detailed calculations of $\Sigma_{12}$ only. Using
Eqs.~(\ref{eq:SL}) and~(\ref{eq:SR}), the exact solution of the
Dyson equation for $\Sigma_{12}$ in Eq.~(\ref{eq:Dyseq}) has the
form,
\begin{align}\label{eq:Sigma12gen}
  \Sigma_{12}(E,\mathbf{q})= & -\frac{1}{4\alpha\varsigma_{1}\varsigma_{2}\mu^{2}(\hat{q}\mathbf{B})^{2}
  +(\mu B)^{2}[\alpha^{2}(1-\varsigma_{1}^{2})(1-\varsigma_{2}^{2})-2\alpha(1+\varsigma_{1}\varsigma_{2})+1]}
  \nonumber
  \\
  & \times
  \big\{
    \alpha(\varsigma_{1}-\varsigma_{2})\mu(\hat{q}\mathbf{B})
    +(\bm{\sigma}\cdot[\mathrm{i}\alpha(\varsigma_{1}+\varsigma_{2})\mu(\hat{q}\times\mathbf{B})
    -2\alpha\varsigma_{1}\varsigma_{2}\mu(\hat{q}\mathbf{B})\hat{q}
    \nonumber
    \\
    & +
    \mu\mathbf{B}(\alpha(1+\varsigma_{1}\varsigma_{2})-1)])
  \big\},
\end{align}
where
\begin{equation}
  \alpha=\frac{4(E-E_{1}+\mathrm{i}0)(E-E_{2}+\mathrm{i}0)}{\mu^{2}B^{2}(1-\varsigma_{1}^{2})(1-\varsigma_{2}^{2})},
\end{equation}
and $E_{1,2}=\sqrt{q^{2}+m_{1,2}^{2}}$. For ultrarelativistic neutrinos,
one has that $\varsigma_{1,2}\to1$, and Eq.~(\ref{eq:Sigma12gen})
takes the form,
\begin{equation}\label{eq:Sigma12ur}
  \Sigma_{12}=-\frac{\mu(\bm{\sigma}\cdot[\mathbf{B}+\mathrm{i}(\hat{q}\times\mathbf{B})-(\hat{q}\mathbf{B})\hat{q}])}
  {2[(E-E_{1}+\mathrm{i}0)(E-E_{2}+\mathrm{i}0)-\mu^{2}B^{2}+\mu^{2}(\hat{q}\mathbf{B})^{2}]}.
\end{equation}
Analogously to Eq.~(\ref{eq:Sigma12ur}), one gets that $\Sigma_{21}=-\Sigma_{12}$.

\section{Transition probability of the spin-flavor precession}\label{sec:TRANSPROB}

In this section, we calculate the 3D Fourier transform of the dressed
propagators, derived in Sec.~\ref{sec:PROPB}, which the matrix element
depends on; cf. Eq.~(\ref{eq:matrelemu}). Then, we compute the transition
probability for the $\nu_{e}\to\bar{\nu}_{\mu}$ precession.

We present the detailed calculation of the $\Sigma_{12}$ contribution
to the matrix element. The magnetic field is taken to be transverse with respect to $\mathbf{L}$, $\mathbf{B}\perp\mathbf{L}$. For instance, we suppose that $\mathbf{L}=L\mathbf{e}_{z}$ and $\mathbf{B}=B\mathbf{e}_{x}$,
where $\mathbf{e}_{x,z}$ are the unit vectors along the $x$- and $z$-axes.
In the QM approach, a neutrino is on the mass shell, i.e. the neutrino momentum is also perpendicular to $\mathbf{B}$. In the QFT treatment, a neutrino is a virtual particle. Thus, $\mathbf{q}$ is arbitrary and, hence, $(\hat{q}\cdot\mathbf{B}) \neq 0$ in Eq.~\eqref{eq:Sigma12ur}.

We should keep the terms (i) $\propto \mu(\bm{\sigma}\cdot\hat{q})(\hat{q}\cdot\mathbf{B})$ and (ii) $\propto \mu^{2}(\hat{q}\cdot\mathbf{B})^{2}$ in the numerator and the denominator of $\Sigma_{12}$ in Eq.~\eqref{eq:Sigma12ur} since this kind of terms is of the same order of magnitude as other terms in the propagator. At least, one has no reasons to neglect the terms (i) and (ii) before the calculation of the matrix element.

To evaluate the contribution of the terms (i) and (ii) to the matrix element we adopt the forward scattering approximation, i.e. we assume that $\mathbf{p}_{e}\upuparrows\mathbf{p}_{\bar{\mu}}\upuparrows\mathbf{L}$ to simplify some integrals computation. This approximation is valid for ultrarelativistic charged leptons. The leptons with arbitrary energies are considered shortly in Sec.~\ref{sec:ARBENER}.

Following Ref.~\cite{Dvo25}, we use the cylindrical
coordinates $(\rho,\phi,z)$ to compute the integral in Eq.~(\ref{eq:matrelemu}),
$\mathbf{q}=\rho(\cos\phi\mathbf{e}_{x}+\sin\phi\mathbf{e}_{y})+z\mathbf{e}_{z}$.
In the forward scattering approximation, the spinors of the charged leptons are $\kappa_{e\mathrm{L}}^{\mathrm{T}}=(0,1)$
and $\kappa_{\bar{\mu}\mathrm{R}}^{\mathrm{T}}=(1,0)$. After averaging the contribution of $\Sigma_{12}$ to the matrix element over $\kappa_{\bar{\mu}\mathrm{R}}$ and $\kappa_{e\mathrm{L}}$, we can drop some terms which vanish in the integration over $\phi$ because of the symmetry properties of the integrand. Therefore, the remainung nontrivial contribution reads
\begin{equation}\label{eq:Sigma12av}
  \kappa_{\bar{\mu}\mathrm{R}}^{\dagger}\Sigma_{12}\kappa_{e\mathrm{L}}= \frac{f}{\mathcal{E}_{+}(\mathbf{q})-\mathcal{E}_{-}(\mathbf{q})}
  \left(
    \frac{1}{E-\mathcal{E}_{+}(\mathbf{q})+\mathrm{i}0}-\frac{1}{E-\mathcal{E}_{-}(\mathbf{q})+\mathrm{i}0}
  \right),
\end{equation}
where
\begin{equation}\label{eq:faux}
  f(\rho,\phi,z)=-\frac{\mu B}{2}
  \left(
    1+\frac{z}{\sqrt{\rho^{2}+z^{2}}}-\frac{\rho^{2}\cos^{2}\phi}{\rho^{2}+z^{2}}
  \right),
\end{equation}
and
\begin{equation}\label{eq:Epmq}
  \mathcal{E}_{\pm}(\mathbf{q})=\bar{E}\pm\sqrt{\Delta E^{2}+(\mu B)^{2}
  \left(
    1-\frac{\rho^{2}\cos^{2}\phi}{\rho^{2}+z^{2}}
  \right)}.
\end{equation}
Here $\bar{E}=[E_{2}(\mathbf{q})+E_{1}(\mathbf{q})]/2$ and $\Delta E=[E_{2}(\mathbf{q})-E_{1}(\mathbf{q})]/2$.

Let us compute one of the integrals, which contributes to Eq.~(\ref{eq:Sigma12av}).
Namely
\begin{align}\label{eq:Ipdef}
  I_{+}= & \frac{1}{(2\pi)^{3}}\int_{0}^{2\pi}\mathrm{d}\phi\int_{0}^{\infty}\rho\mathrm{d}\rho\int_{-\infty}^{+\infty}\mathrm{d}z
  \frac{fe^{\mathrm{i}zL}}{(\mathcal{E}_{+}-\mathcal{E}_{-})(E-\mathcal{E}_{+}+\mathrm{i}0)}
  \nonumber
  \\
  & =
  -\frac{\mathrm{i}}{(2\pi)^{2}}\int_{0}^{2\pi}\mathrm{d}\phi
  \sum_{ \left\{ z_{+}\right\} }
  \int_{0}^{\infty}\rho\mathrm{d}\rho
  \left.
    \frac{f(\rho,\phi,z_{+})e^{\mathrm{i}z_{+}L}}{(\mathcal{E}_{+}-\mathcal{E}_{-})\frac{\mathrm{d}\mathcal{E}_{+}}{\mathrm{d}z}}
  \right|_{z=z_{+}},
\end{align}
where $z_{+}(\rho,\phi)$ are the roots of the equation $E-\mathcal{E}_{+}(\rho,\phi,z)+\mathrm{i}0=0$.
These roots are in the upper half-plane. Since we take that $L>0$,
we close the contour in the upper half-plane while integrating over
$z$.

We define the new quantity
\begin{equation}\label{eq:EpmE}
  \mathcal{E}_{+}(E)=E+\sqrt{\left(\frac{\Delta m^{2}}{4E}\right)^{2}+(\mu B)^{2}\left(1-\frac{\rho^{2}}{E^{2}}\cos^{2}\phi\right)},
\end{equation}
where $\Delta m^{2}=m_{2}^{2}-m_{1}^{2}>0$. Note that $\mathcal{E}_{+}(E)\neq\mathcal{E}_{+}(\mathbf{q})$;
cf. Eq.~(\ref{eq:Epmq}). We can approximately set that $\left.\frac{\mathrm{d}\mathcal{E}_{+}(\mathbf{q})}{\mathrm{d}z}\right|{}_{z=z_{+}}\approx\frac{z_{+}}{\mathcal{E}_{+}(E)}$
and $\mathcal{E}_{+}^{2}(E)\approx\rho^{2}+z_{+}^{2}$. Moreover,
for ultrarelativistic neutrinos, having $\mathfrak{E} \ll E$,
and a rather weak magnetic field, obeying $(\mu B)^{2}\ll E\mathfrak{E}$,
one has that
\begin{equation}\label{eq:EpEp0}
  \mathcal{E}_{+}(E)\approx
  E+\mathfrak{E}
  \left(
    1-\frac{(\mu B)^{2}\rho^{2}\cos^{2}\phi}{2E^{2}\mathfrak{E}^{2}}
  \right),
\end{equation}
where
\begin{equation}\label{eq:Ep0}
  \mathfrak{E}=\sqrt{(\mu B)^{2}+\left(\frac{\Delta m^{2}}{4E}\right)^{2}}.
\end{equation}
Using Eq.~(\ref{eq:EpEp0}), one gets the following expressions for the poles in the
explicit form:
\begin{align}\label{eq:zpexpl}
  z_{+} & =\sqrt{1+\frac{(\mu B)^{2}}{E\mathfrak{E}}\cos^{2}\phi}\times
  \begin{cases}
    \sqrt{\rho_{0}^{2}-\rho^{2}} + \mathrm{i}0, & \text{if}\ \rho<\rho_{0},\\
    \mathrm{i}\sqrt{\rho^{2}-\rho_{0}^{2}}, & \text{if}\ \rho>\rho_{0},
  \end{cases}
\end{align}
where $\rho_{0}=(E+\mathfrak{E})/\sqrt{1+\frac{(\mu B)^{2}}{E\mathfrak{E}}\cos^{2}\phi}$.

Using Eqs.~(\ref{eq:faux}) and~(\ref{eq:zpexpl}), we rewrite Eq.~(\ref{eq:Ipdef})
as
\begin{align}\label{eq:Ipphirho}
  I_{+}= & \frac{\mathrm{i}\mu B(E+\mathfrak{E})}{16\pi^{2}\mathfrak{E}}
  \int_{0}^{2\pi}\frac{\mathrm{d}\phi}{\sqrt{1+\frac{(\mu B)^{2}}{E\mathfrak{E}}\cos^{2}\phi}}
  \Bigg[
    \int_{0}^{\rho_{0}}\frac{\rho\mathrm{d}\rho
    \exp
    \left(
      \mathrm{i}\tilde{L}\sqrt{\rho_{0}^{2}-\rho^{2}}
    \right)}{\sqrt{\rho_{0}^{2}-\rho^{2}}}
    \nonumber
    \\
    & \times
    \left(
      1-\rho^{2}\cos^{2}\phi
      \left(\frac{1}{(E+\mathfrak{E})^{2}}
       -\frac{(\mu B)^{2}}{2E^{2}\mathfrak{E}^{2}}
      \right)
      +\frac{\sqrt{\rho_{0}^{2}-\rho^{2}}}{\tilde{E}
      \left(
        1+\frac{\mathfrak{E}}{E}
      \right)}
    \right)
    \nonumber
    \\
    & -
    \mathrm{i}\int_{\rho_{0}}^{\infty}\frac{\rho\mathrm{d}\rho
    \exp
    \left(
      -\tilde{L}\sqrt{\rho^{2}-\rho_{0}^{2}}
    \right)}{\sqrt{\rho^{2}-\rho_{0}^{2}}}
    \notag
    \\
    & \times
    \left(
      1-\rho^{2}\cos^{2}\phi
      \left(\frac{1}{(E+\mathfrak{E})^{2}}
       -\frac{(\mu B)^{2}}{2E^{2}\mathfrak{E}^{2}}
      \right)
      +\frac{\mathrm{i}\sqrt{\rho^{2} - \rho_{0}^{2}}}{\tilde{E}
      \left(
        1+\frac{\mathfrak{E}}{E}
      \right)}
    \right)
  \Bigg],
\end{align}
where
\begin{equation}\label{eq:Ipfin}
  \tilde{L}=L\sqrt{1+\frac{(\mu B)^{2}}{E\mathfrak{E}}\cos^{2}\phi},
  \quad
  \tilde{E}=\frac{E}{\sqrt{1+\frac{(\mu B)^{2}}{E\mathfrak{E}}\cos^{2}\phi}}.
\end{equation}
Making the integrations over $\rho$, we represent Eq.~(\ref{eq:Ipphirho})
in the form,
\begin{align}\label{eq:Ipapp}
  I_{+}\approx & \frac{\mu B(E+\mathfrak{E})
  e^{\mathrm{i}(E+\mathfrak{E})}}{8\pi{}^{2}\mathfrak{E}L}
  \int_{0}^{2\pi}\frac{\mathrm{d}\phi}{1+\frac{(\mu B)^{2}}{E\mathfrak{E}}\cos^{2}\phi}
  \nonumber
  \\
  & \approx
  \frac{\mu B(E+\mathfrak{E})
  e^{\mathrm{i}(E+\mathfrak{E})L}}{4\pi\mathfrak{E}L}
  \left(
    1-\frac{(\mu B)^{2}}{2E\mathfrak{E}}
  \right),
\end{align}
where we assume that the distance between the source and the detector
is great, $L\gg E^{-1}$.

Analogously to Eq.~(\ref{eq:Ipdef}), one computes the integral,
\begin{equation}\label{eq:Imdef}
  I_{-}=\int\frac{\mathrm{d}^{3}q}{(2\pi)^{3}}
  \frac{fe^{\mathrm{i}\mathbf{qL}}}{(\mathcal{E}_{+}-\mathcal{E}_{-})(E-\mathcal{E}_{-}+\mathrm{i}0)}.
\end{equation}
We give just the final result for $I_{-}$ in Eq.~(\ref{eq:Imdef}),
\begin{equation}\label{Imfin}
  I_{-}\approx
  \frac{\mu B(E-\mathfrak{E})
  e^{\mathrm{i}(E-\mathfrak{E})L}}{4\pi\mathfrak{E}L}
  \left(
    1-\frac{(\mu B)^{2}}{2E\mathfrak{E}}
  \right).
\end{equation}
As we mentioned in Sec.~\ref{sec:PROPB}, the contributions of the diagonal propagators $\Sigma_{aa}$
to the matrix element $\mathcal{M}_{e\to\bar{\mu}}$ are vanishing.

Since $\Sigma_{21}=-\Sigma_{12}$, one gets from Eqs.~(\ref{eq:matrelemu}),
(\ref{eq:Ipfin}), and~(\ref{Imfin}) that
\begin{equation}\label{eq:matrelfin}
  \mathcal{M}_{e\to\bar{\mu}}=I_{-}-I_{+}=
  -\frac{\mathrm{i} E e^{\mathrm{i}EL}}{2\pi L}
  \frac{\mu B}{\mathfrak{E}}
  \left(
    1-\frac{(\mu B)^{2}}{2\mathfrak{E}E}
  \right)
  \left(
    \sin\mathfrak{E}L
    - \mathrm{i} \frac{\mathfrak{E}}{E}\cos\mathfrak{E}L
  \right).
\end{equation}
The probability of the process $\nu_{e}\to\bar{\nu}_{\mu}$ is $P_{\nu_{e}\to\bar{\nu}_{\mu}}\propto|\mathcal{M}_{e\to\bar{\mu}}|^{2}$.
Based on Eq.~(\ref{eq:matrelfin}),
we obtain that
\begin{equation}\label{eq:trprob}
  P_{\nu_{e}\to\bar{\nu}_{\mu}}(L)
  \approx 
  P_\mathrm{max}
  \sin^{2}
  \left(
    \sqrt{(\mu B)^{2}+\left(\frac{\Delta m^{2}}{4E}\right)^{2}}L
  \right),
\end{equation}
where the amplitude of the transition probability has two terms,  $P_\mathrm{max} = P_\mathrm{max}^{(\mathrm{qm})} + P_\mathrm{max}^{(\mathrm{qft})}$. These terms have the following forms:
\begin{equation}\label{eq:Ptrqft0}
  P_\mathrm{max}^{(\mathrm{qm})} =
  \frac{(\mu B)^{2}}{(\mu B)^{2}+\left(\frac{\Delta m^{2}}{4E}\right)^{2}},
  \quad
  P_\mathrm{max}^{(\mathrm{qft})} =
  - P_\mathrm{max}^{(\mathrm{qm})} \frac{(\mu B)^{2}}
  {E \sqrt{ (\mu B)^{2}+\left(\frac{\Delta m^{2}}{4E}\right)^{2} } }.
\end{equation}
We shall see shortly in Sec.~\ref{sec:SFOQM} that $P_\mathrm{max}^{(\mathrm{qm})}$ in Eq.~\eqref{eq:Ptrqft0} reproduces the result of the QM approach. The term $P_\mathrm{max}^{(\mathrm{qft})}$ in Eq.~\eqref{eq:Ptrqft0} is the QFT contribution to the transition probability. It arises from the terms (i) and (ii) in the propagator $\Sigma_{12}$

To derive Eqs.~\eqref{eq:trprob} and~\eqref{eq:Ptrqft0}, we take into account Eq.~(\ref{eq:Ep0}) and assume that $\mu B \sim \mathfrak{E}$. Moreover, we neglect the terms $\sim (\mathfrak{E}/E)^2 \ll 1$, whereas we keep the terms $\sim \mu B / E$ or $\sim \mathfrak{E}/E$.

One can see in Eq.~\eqref{eq:Ptrqft0} that the QFT contribution to the transition probability is negative. Moreover, $|P_\mathrm{max}^{(\mathrm{qft})}|/P_\mathrm{max}^{(\mathrm{qm})} \sim \mu B / E \sim \mathfrak{E}/E$. Thus, the QFT term is much smaller than the QM one for all realistic oscillations channels where neutrinos are ultrarelativistic.

The latter result proves the assumption that the terms (i) and (ii) in $\Sigma_{12}$ in Eq.~\eqref{eq:Sigma12ur} can be safely neglected for ultrarelativistic neutrinos since their contribution to the transition probability is small. We recall that all the terms in $\Sigma_{12}$ in Eq.~\eqref{eq:Sigma12ur}, including (i) and (ii), are of the same order of magnitude. Therefore, before the transition probability computation, one cannot say anything about the magnitude of certain terms in $\Sigma_{12}$.

\subsection{Scattering of charged leptons with arbitrary energies}
\label{sec:ARBENER}

We have just demonstrated that the contributions of the terms (i) and (ii) in the propagator in Eq.~\eqref{eq:Sigma12ur}, which have the QFT origin, to the transition probability of the spin-flavor precession are small.
The washing out of these terms happens for ultrarelativistic neutrinos. That is why, we can neglect the terms (i) and (ii) while we study the case of charged leptons with arbitrary energies, with neutrinos being ultrarelativistic.

For simplicity, we take that an incoming electron still propagates along $\mathbf{L}$. The outgoing antimuon can have an arbitrary direction of its momentum. The spin of an electron is taken to be opposite to the electron momentum, $\mathbf{s}_e \downuparrows \mathbf{p}_e$, whereas the antimuon spin is supposed to be along the particle momentum, $\mathbf{s}_{\bar{\mu}} \upuparrows \mathbf{p}_{\bar{\mu}}$.

The chiral projections of leptons bispinors have the forms,
\begin{equation}
  \kappa_{e\mathrm{L}} =
  \sqrt{\frac{1+v_e}{2}} 
  \left(
    \begin{array}{c}
      0\\
      1
    \end{array}
  \right),
  \quad
  \kappa_{\bar{\mu}\mathrm{R}} =
  -\sqrt{\frac{1+v_{\bar{\mu}}}{2}} 
  \left(
    \begin{array}{c}
      e^{-\mathrm{i}\phi_{\bar{\mu}}} \cos \frac{\theta_{\bar{\mu}}}{2} \\
      e^{\mathrm{i}\phi_{\bar{\mu}}} \sin \frac{\theta_{\bar{\mu}}}{2}
    \end{array}
  \right),
\end{equation}
where $v_e$ and $v_{\bar{\mu}}$ are the velocities of the electron and the antimuon, and the spherical angles $\phi_{\bar{\mu}}$ and $\theta_{\bar{\mu}}$ fix the direction of the antimuon momentum with respect to $\mathbf{e}_z$.

The averaging of $\Sigma_{12}$ over $\kappa_{\bar{\mu}\mathrm{R}}$ and $\kappa_{e\mathrm{L}}$ is similar to that in Eq.~\eqref{eq:Sigma12av} except the following replacements:
\begin{align}\label{eq:fEpm}
  f(\rho,\phi,z) \to &
  \sqrt{(1+v_e)(1+v_{\bar{\mu}})}  
  e^{\mathrm{i}\phi_{\bar{\mu}}} \cos
  \left(
    \frac{\theta_{\bar{\mu}}}{2}
  \right)
  \frac{\mu B}{4}
  \left(
    1+\frac{z}{\sqrt{\rho^{2}+z^{2}}}
  \right),
  \notag
  \\
  \mathcal{E}_{\pm}(\mathbf{q}) \to &
  \bar{E}\pm\sqrt{\Delta E^{2}+(\mu B)^{2}}.
\end{align}
The subsequent computation of the matrix element is identical to that in Sec.~\ref{sec:TRANSPROB}. The final result reads
\begin{equation}\label{eq:matrelarben}
  \mathcal{M}_{e\to\bar{\mu}}=
  \frac{\mathrm{i} E e^{\mathrm{i}EL}}{4\pi L}
  \sqrt{(1+v_e)(1+v_{\bar{\mu}})}
  e^{\mathrm{i}\phi_{\bar{\mu}}} \cos
  \left(
    \frac{\theta_{\bar{\mu}}}{2}
  \right)
  \frac{\mu B}{\mathfrak{E}}
  \sin\mathfrak{E}L,
\end{equation}
i.e. we remove the QFT corrections from Eq.~\eqref{eq:matrelfin} and take into account the additional factor, $-e^{\mathrm{i}\phi_{\bar{\mu}}} \cos \left( \tfrac{\theta_{\bar{\mu}}}{2} \right) \sqrt{(1+v_e)(1+v_{\bar{\mu}})}/2$.

The measurable quantity of the process $e+N \to \bar{\mu} + \tilde{N}'$ is the differential cross-section~\cite[p.~252]{BerLifPit82},
\begin{equation}\label{eq:csdef}
  \mathrm{d}\sigma_{e\to\bar{\mu}} = G_\mathrm{int}^4 E^2 
  |\mathcal{M}_{e\to\bar{\mu}}|^2
  \frac{v_{\bar{\mu}}}{v_e}
  \mathrm{d}\Omega_{\bar{\mu}},
\end{equation}
where $\mathrm{d}\Omega_{\bar{\mu}} = 2\pi \sin\theta_{\bar{\mu}} \mathrm{d}\theta_{\bar{\mu}}$ is the solid angle of the antimuon momentum space.

Integrating Eq.~\eqref{eq:csdef} over $\mathrm{d}\Omega_{\bar{\mu}}$, one gets that the total cross-section is
\begin{equation}\label{eq:totcs}
  \sigma_{e\to\bar{\mu}} = \int \mathrm{d}\sigma_{e\to\bar{\mu}}
  =
  \frac{E^4 G_\mathrm{int}^4}{2\pi L^2}
  P^{(\mathrm{eff})}_{\nu_e \to \bar{\nu}_\mu},
  \quad
  P^{(\mathrm{eff})}_{\nu_e \to \bar{\nu}_\mu} =
  (1+v_e)(1+v_{\bar{\mu}}) 
  \frac{v_{\bar{\mu}}}{4v_e}
  \left(
    \frac{\mu B}{\mathfrak{E}}
  \right)^2
  \sin^2 (\mathfrak{E} L).
\end{equation}
%
One can see in Eq.~\eqref{eq:totcs} that the total cross-section oscillates in space and is proportional to the QM transition probability, $P_\mathrm{max}^{(\mathrm{qm})} \sin^2 (\mathfrak{E} L)$; cf. Eqs.~\eqref{eq:trprob} and~\eqref{eq:Ptrqft0}.

For ultrarelativistic charged leptons, $P^{(\mathrm{eff})}_{\nu_e \to \bar{\nu}_\mu} \approx P^{(\mathrm{qm})}_{\nu_e \to \bar{\nu}_\mu}$ since $v_{\bar{\mu}} \sim v_e \sim 1$. In general situation, $P^{(\mathrm{eff})}_{\nu_e \to \bar{\nu}_\mu} < P^{(\mathrm{qm})}_{\nu_e \to \bar{\nu}_\mu}$ since $\mathfrak{m}_\mu > \mathfrak{m}_e$. Thus, the factor $(1+v_e)(1+v_{\bar{\mu}})v_{\bar{\mu}} / (4v_e) < 1$ should be accounted for while designing a neutrino oscillations experiment.

\section{Quantum mechanical description of the spin-flavor precession}\label{sec:SFOQM}

In this section, we analyze the QM description of
the spin-flavor precession of Majorana neutrinos in an external magnetic
field. This study can be done in both flavor and mass bases. We take
the flavor basis since it is more common in the standard QM
approach.

We suppose that two flavor neutrinos, $\nu_{e}$ and $\nu_{\mu}$,
propagate along the $z$-axis and interact with the transverse external
magnetic field $\mathbf{B}=(B,0,0)$, as in Sec.~\ref{sec:PROPB}.
Despite one can reveal the nature of neutrinos only in the mass basis,
we can still assume that $\nu_{\lambda}$, $\lambda=e,\mu$, are Majorana
particles. The interaction of these neutrinos with the magnetic field
is owing to the transition magnetic moment $\mu$. The effective Schr\"odinger
equation in the basis $\nu^{\mathrm{T}}=(\nu_{e},\nu_{\mu},\bar{\nu}_{e},\bar{\nu}_{\mu})$
was obtained in Ref.~\cite{LimMar88},
\begin{equation}\label{eq:Scheqfl}
  \mathrm{i}\dot{\nu}=H_{f}\nu,
  \quad
  H_{f}=
  \left(
    \begin{array}{cccc}
      -\frac{\Delta m^{2}}{4E_{\nu}}\cos2\theta & \frac{\Delta m^{2}}{4E_{\nu}}\sin2\theta & 0 & \mathrm{i}\mu B \\
      \frac{\Delta m^{2}}{4E_{\nu}}\sin2\theta & \frac{\Delta m^{2}}{4E_{\nu}}\cos2\theta & -\mathrm{i}\mu B & 0 \\
      0 & \mathrm{i}\mu B & -\frac{\Delta m^{2}}{4E_{\nu}}\cos2\theta & \frac{\Delta m^{2}}{4E_{\nu}}\sin2\theta \\
      -\mathrm{i}\mu B & 0 & \frac{\Delta m^{2}}{4E_{\nu}}\sin2\theta & \frac{\Delta m^{2}}{4E_{\nu}}\cos2\theta
    \end{array}
  \right),
\end{equation}
where $E_{\nu}$ is the mean energy of neutrinos, and $\theta$ is
the vacuum mixing angle defined in Eq.~\eqref{eq:flmassrel}. We already
mentioned in Sec.~\ref{sec:ELECTRMAJ} (see also Ref.~\cite{Pas00})
that the magnetic moments matrix $(\mu_{ab})=\mu\sigma_{2}$ is hermitian
and purely imaginary in the mass basis. Taking into account Eq.~\eqref{eq:flmassrel},
we get that this matrix in the flavor basis reads $(M_{\lambda\lambda'})=U(\mu_{ab})U^{\dagger}=\mu\sigma_{2}$,
that is used in Eq.~(\ref{eq:Scheqfl}).

The general solution of Eq.~(\ref{eq:Scheqfl}), which satisfies
the given initial condition $\nu_{0}$, has the form,
\begin{equation}\label{eq:gensolfl}
  \nu(t)=
  \left(
    \left[
      \left(
        U_{1}\otimes U_{1}^{\dagger}
      \right)+
      \left(
        U_{2}\otimes U_{2}^{\dagger}
      \right)
    \right]
    e^{-\mathrm{i}\mathfrak{\mathfrak{E}_{\nu}}t}+
    \left[
      \left(
        V_{1}\otimes V_{1}^{\dagger}
      \right)+
      \left(
        V_{2}\otimes V_{2}^{\dagger}
      \right)
    \right]
    e^{\mathrm{i}\mathfrak{E}_{\nu}t}
  \right)
  \nu_{0},
\end{equation}
where
\begin{equation}
  \mathfrak{E}_{\nu}=\sqrt{\left(\frac{\Delta m^{2}}{4E_{\nu}}\right)^{2}+(\mu B)^{2}},
\end{equation}
and
\begin{align}\label{eq:U12V12}
  U_{1} & =\frac{1}{\sqrt{2\mathfrak{E}_{\nu}}}
  \left(
    \begin{array}{c}
      \frac{\mathrm{i}\mu B}{\sqrt{\mathfrak{E}_{\nu}+\frac{\Delta m^{2}}{4E_{\nu}}\cos2\theta}} \\
      0 \\
      \frac{\frac{\Delta m^{2}}{4E_{\nu}}\sin2\theta}{\sqrt{\mathfrak{E}_{\nu}+\frac{\Delta m^{2}}{4E_{\nu}}\cos2\theta}} \\
      \sqrt{\mathfrak{E}_{\nu}+\frac{\Delta m^{2}}{4E_{\nu}}\cos2\theta}
    \end{array}\right),
    \quad
    V_{1}=\frac{1}{\sqrt{2\mathfrak{E}_{\nu}}}
    \left(
      \begin{array}{c}
        \sqrt{\mathfrak{E}_{\nu}+\frac{\Delta m^{2}}{4E_{\nu}}\cos2\theta} \\
        -\frac{\frac{\Delta m^{2}}{4E_{\nu}}\sin2\theta}{\sqrt{\mathfrak{E}_{\nu}+\frac{\Delta m^{2}}{4E_{\nu}}\cos2\theta}} \\
        0 \\
        \frac{\mathrm{i}\mu B}{\sqrt{\mathfrak{E}_{\nu}+\frac{\Delta m^{2}}{4E_{\nu}}\cos2\theta}}
     \end{array}
   \right),
   \nonumber
   \\
  U_{2} & =\frac{1}{\sqrt{2\mathfrak{E}_{\nu}}}
  \left(
    \begin{array}{c}
      \frac{\frac{\Delta m^{2}}{4E_{\nu}}\sin2\theta}{\sqrt{\mathfrak{E}_{\nu}+\frac{\Delta m^{2}}{4E_{\nu}}\cos2\theta}} \\
      \sqrt{\mathfrak{E}_{\nu}+\frac{\Delta m^{2}}{4E_{\nu}}\cos2\theta} \\
      \frac{\mathrm{i}\mu B}{\sqrt{\mathfrak{E}_{\nu}+\frac{\Delta m^{2}}{4E_{\nu}}\cos2\theta}} \\
      0
    \end{array}
  \right),
  \quad
  V_{2}=\frac{1}{\sqrt{2\mathfrak{E}_{\nu}}}
  \left(
    \begin{array}{c}
      0 \\
      \frac{\mathrm{i}\mu B}{\sqrt{\mathfrak{E}_{\nu}+\frac{\Delta m^{2}}{4E_{\nu}}\cos2\theta}} \\
      \sqrt{\mathfrak{E}_{\nu}+\frac{\Delta m^{2}}{4E_{\nu}}\cos2\theta} \\
      -\frac{\frac{\Delta m^{2}}{4E_{\nu}}\sin2\theta}{\sqrt{\mathfrak{E}_{\nu}+\frac{\Delta m^{2}}{4E_{\nu}}\cos2\theta}}
    \end{array}
  \right),
\end{align}
are the eigenvectors which are normalized, $|U_{1,2}|^{2}=|V_{1,2}|^{2}=1$.

We suppose that $\nu(0)\equiv\nu_{e}$ initially. Thus, we take $\nu_{0}^{\mathrm{T}}=(1,0,0,0)$
in Eq.~(\ref{eq:gensolfl}). Using Eq.~(\ref{eq:U12V12}), the general
solution in Eq.~(\ref{eq:gensolfl}), satisfying the given initial
condition, takes the form,
\begin{equation}\label{eq:gensolinicond}
  \nu(t)=\frac{1}{2\mathfrak{E}_{\nu}}
  \left(
    \begin{array}{c}
      \frac{(\mu B)^{2}+
      \left(
        \frac{\Delta m^{2}}{4E_{\nu}}
      \right)^{2}
      \sin^{2}2\theta}{\mathfrak{E}_{\nu}+\frac{\Delta m^{2}}{4E_{\nu}}\cos2\theta}e^{-\mathrm{i}\mathfrak{E}_{\nu}t}+
      \left(
        \mathfrak{E}_{\nu}+\frac{\Delta m^{2}}{4E_{\nu}}\cos2\theta
      \right)
      e^{\mathrm{i}\mathfrak{E}_{\nu}t} \\
      -\mathrm{i}\frac{\Delta m^{2}}{2E_{\nu}}\sin2\theta\sin\mathfrak{E}_{\nu}t \\
      0 \\
      -2\mu B\sin\mathfrak{E}_{\nu}t
    \end{array}
  \right),
\end{equation}
Based on Eq.~(\ref{eq:gensolinicond}), we get that $\bar{\nu}_{\mu}(t)=-\frac{\mu B}{\mathfrak{E}_{\nu}}\sin\mathfrak{E}_{\nu}t$.
The transition probability for the process $\nu_{e}\to\bar{\nu}_{\mu}$
is
\begin{equation}\label{eq:Ptrqm}
  P_{\nu_{e}\to\bar{\nu}_{\mu}}(t)=|\bar{\nu}_{\mu}(t)|^{2}=
  \frac{(\mu B)^{2}}{(\mu B)^{2}+
  \left(
    \frac{\Delta m^{2}}{4E_{\nu}}
  \right)^{2}}\sin^{2}
  \left(
    \sqrt{(\mu B)^{2}+
    \left(
      \frac{\Delta m^{2}}{4E_{\nu}}
    \right)^{2}}t
  \right).
\end{equation}
It is interesting to mention that the transition probability in Eq.~(\ref{eq:Ptrqm})
does not depend on the vacuum mixing angle.

Now, we can compare the QM and QFT
approaches to the neutrino spin-flavor precession. If neutrinos
are ultrareletivistic, we can put $E_{\nu}=E$, $\mathfrak{E}_{\nu}=\mathfrak{E}$,
and $t\approx L$ in Eq.~(\ref{eq:Ptrqm}). Then, we omit the small
correction $P_\mathrm{max}^{(\mathrm{qft})}$ in Eq.~\eqref{eq:Ptrqft0}.
Under these approximations, the transition probability in Eqs.~\eqref{eq:trprob}
and~\eqref{eq:Ptrqft0}
coincides with that in Eq.~(\ref{eq:Ptrqm}),
i.e. both approaches provide the same description of the spin-flavor precession.

\section{Discussion}\label{sec:CONCL}

In the present work, we have developed the QFT description of the neutrino
spin-flavor precession in a magnetic field. We were based on the
QFT formalism, which was originally developed in Refs.~\cite{Kob82,GiuKimLee93,GriSto96},
where neutrinos are virtual particles. This method was used in Ref.~\cite{Kob82}
to study neutrino-to-antineutrino oscillations in vacuum. If one deals
with Majorana neutrinos with transition magnetic moments, $\nu\leftrightarrow\bar{\nu}$
oscillations are analogous to spin-flavor transitions in a magnetic
field. We have adapted the results of Ref.~\cite{Kob82} for the
description of the spin-flavor precession.

In Sec.~\ref{sec:ELECTRMAJ}, we have reminded how Majorana neutrinos
can interact with electromagnetic fields. Considering the system of
two neutrinos, we have obtained the wave equations for neutrino mass
eigenstates in an external magnetic field. The magnetic interaction
was shown to mix different mass eigenstates, as well as induce particle-to-antiparticle
transitions.

The basics of the QFT approach for the neutrino spin-flavor precession
have been discussed in Sec.~\ref{sec:QFT}. We have obtained the
matrix element corresponding to the transition $\nu_{\beta}\to\bar{\nu}_{\alpha}$
in an external magnetic field. The dressed propagators for the neutrino
mass eigenstates, accounting for the magnetic interaction, have been
derived in Sec.~\ref{sec:PROPB} for neutrinos with arbitrary energies.
Then, we have considered the case of ultrarelativistic particles.

Finally, we have derived the transition probability for the $\nu_{e}\to\bar{\nu}_{\mu}$ precession in a magnetic field in Sec.~\ref{sec:TRANSPROB} within
the QFT approach. This transition probability has the leading term,
which was shown in Sec.~\ref{sec:SFOQM} to coincide with the result
of the standard QM approach. The coincidence of the predictions of QFT and QM approaches takes place for ultrarelativistic neutrinos. In this situation, we identify the distance between the source and the detector with the propagation time, $L \approx t$. Additionally, one can neglect the QFT contribution to the transition probability for ultrarelativistic neutrinos.

Moreover, we have obtained that the dressed propagator in Eq.~\eqref{eq:Sigma12ur} acquires the terms (i) and (ii) which have the QFT nature. These terms cannot be neglected before the transition probability calculation. We have demonstrated in Sec.~\ref{sec:TRANSPROB} that the QFT terms in the propagator result in 
the correction to the leading term in the transition probability which is small if neutrinos are ultrarelativistic; cf. Eqs.~\eqref{eq:trprob} and~\eqref{eq:Ptrqft0}.

Despite we treat neutrinos as virtual particles, the consideration of the QFT correction to the transition probability shows that the virtuality of neutrinos is negligible. Indeed, in Sec.~\ref{sec:TRANSPROB}, we have demonstrated that the terms (i) and (ii) in $\Sigma_{12}$ can be omitted in a realistic situation. Hence, virtual neutrinos behave as if they were on the mass shell and were described within the QM formalism. In particular, it happens since we study macroscopically large propagation distances.
 
In Sec.~\ref{sec:ARBENER}, we have discussed a realistic situation when charged leptons have arbitrary energies. The forward scattering approximation is not valid in this case. We have derived the effective transition probability in Eq.~\eqref{eq:totcs} which depends on the velocities of incoming and outgoing charged leptons. It turns out that $P^{(\mathrm{eff})}_{\nu_e \to \bar{\nu}_\mu} < P^{(\mathrm{qm})}_{\nu_e \to \bar{\nu}_\mu}$.

The comprehensive application of QFT to the description of neutrino oscillations and, in particular, to the spin-flavor precession, should involve various loop effects along with the diagrams where neutrinos participate on a tree level. We have accounted for some of these loop contributions indirectly. For example, we have assumed that the neutrino magnetic moment already has a certain value instead of calculating the neutrino electromagnetic vertex function (see, e.g., Ref.~\cite{GiuStu15}) at any interaction of the undressed neutrino propagators $S_{a\mathrm{L,R}}$ with the external magnetic potentials $\pm \mathrm{i}\mu (\bm{\sigma}\mathbf{B})$ in Fig.~\ref{fig:Dyson}. The same argument is addressed to the neutrino masses $m_{1,2}$ which are supposed to have physical values. Here, we assume that the model underlying the neutrino mass generation is renormalizable.

Another group of vacuum contributions to neutrino oscillations appears in the approach~\cite{BlaVit96}, where the Fock space of neutrino flavor eigenstates was constructed. It was obtained that a rapidly oscillating term in the transition probability arises along with the conventional slowly varying Pontecorvo contribution~\cite{Bla23,BlaSma25}. This new contribution, resulting from the nontrivial flavor vacuum structure, was shown to vanish for utraretivistic neutrinos. Note that analogous point of view to neutrino oscillations was suggested in Ref.~\cite{Lob19}.

The formalism developed in Refs.~\cite{BlaVit96,Bla23,BlaSma25} describes transitions between different neutrino flavors happening in time. In practice, this kind of transitions can be implemented if one has a permanent source of neutrinos (see also Ref.~\cite{Dvo11}). Note that the diagrammatic approach to the evolution of the interacting fields in a finite time interval was elaborated in Ref.~\cite{Ans23}.

Additionally, the approach, where the mixing between different neutrino types was considered as the perturbation, was developed in Ref.~\cite{BlaSma25}. It is interesting to mention that the mixing was accounted for in Ref.~\cite{BlaSma25} by solving the Dyson equation (see also Ref.~\cite{Tur23}) analogously to our approach where we use the Dyson like equations to take into account the external magnetic field.

In our work, we follow the QFT approach to neutrino oscillations based on the conventional Fock states of the neutrino mass eigenstates~\cite{Kob82,GiuKimLee93,GriSto96} (see, e.g., Eq.~\eqref{eq:propx} where the dressed propagators are defined). The advantage of this formalism consists in the fact one does not need to bother about the Poincar\'e symmetry properties of the Fock states, including the vacuum one; cf. Refs.~\cite{BlaSma25,Lob19}. We also mention that our analysis does not reveal rapidly oscillating vacuum terms in the transition probability predicted in Refs.~\cite{Bla23,BlaSma25}.

One of the main goals of the present work is methodological. It is aimed to
demonstrate that the neutrino spin-flavor precession can be described
purely with help of the standard QFT methods if reasonable approximation
are made.

The major assumption is the ultrarelativity of particles
involved in the system. First, in Sec.~\ref{sec:TRANSPROB}, we assume that charged leptons are utrarelativistic, i.e. $E_{\alpha,\beta} \gg \mathfrak{m}_{\alpha,\beta}$. 
The assumption of ultrarelativistic charged leptons was used to derive the transition probability for the spin-flavor precession in the forward scattering approximation. We considered a general situation of charged leptons with arbitrary energies in Sec.~\ref{sec:ARBENER}. Thus, the assumption of ultrarelativistic charged leptons is not crucial for the results of the work.

The most important approximation is the ultrarelativity of virtual neutrinos. It can be expressed in the form, $E = (E_{\alpha} + E_{\beta})/2 \gg m_{1,2}$. Along with the spin-flavor precession under the influence of a magnetic field, for Majorana particles, one has neutrino-to-antineutrino transitions owing to a nonzero neutrino mass. Thus, the series in Eq.~\eqref{eq:Dysser}, would have a more complicated form. Moreover, the bare propagators, derived in Appendix, have rather simple form for ultrarelativistic neutrinos. Finally, without the assumption of the neutrinos ultrarelativity, the computation of the contour integrals, $I_+$ and $I_-$, would be impossible in Sec.~\ref{sec:TRANSPROB}. For neutrinos with arbitrary energies, the equation $E-\mathcal{E}_{\pm}(\mathbf{q})+\mathrm{i}0=0$, which defines the poles, has eight roots which should be found analytically. Thus, the assumption of ultrarelativistic neutrinos in crucial to treat the spin-flavor precession in frames of QFT.

The next important
approximation, which was also mentioned in Ref.~\cite{Dvo25}, is the
consideration of two mass eigenstates only. This feature allows one
to avoid the branching of the Feynman graphs in Fig.~\ref{fig:Dyson}.

In deriving of Eq.~\eqref{eq:Ipapp}, we made an assumption of the great propagation distance $L$. It allows one to avoid terms $\propto L^{-2}$ in Eq.~\eqref{eq:Ipapp}. It is, however, known that the coherence in neutrino oscillations is lost if $L > L_\mathrm{coh}$, where $L_\mathrm{coh}$ is the coherence length. The decoherence effects in neutrino oscillations in vacuum in frames of both QM and QFT approaches have been studied in Refs.~\cite{NauNau20,AkhKop10}.

In fact, numerous factors can contribute to the coherence length. For instance, it depends on a particular experiment where neutrino oscillations are studied (see, e.g., Ref.~\cite{GriStoMoh99}). A careful determination of $L_\mathrm{coh}$ requires the replacement of the plane waves wavefunctions of charged leptons with wave packets of a finite width $\sigma_x$~\cite{GiuKimLee93,GriStoMoh99}. This approach is applicable for the studies of neutrino oscillations in vacuum in frames of QFT. In that situation, the expression for the matrix element can be derived analytically for plane waves wavefunctions of charged leptons, with both neutrinos and charged leptons having arbitrary energies.

In our case, when we study neutrinos interacting with an external magnetic field, the integrations in Sec.~\ref{sec:TRANSPROB} are quite nontrivial even for plane waves charged leptons. As we mentioned earlier, the matrix element can be derived only in the approximation of ultrarelativistic neutrinos. Therefore, a careful consideration of the decoherence in the spin-flavor precession in frames of QFT is a quite complicated task which is beyond the scope of the present work.

We can only say that the coherence length is infinite in our case~\cite{GiuKimLee93}. It results from the fact that we use the plane wave approximation for charged leptons wavefunctions in Eq.~\eqref{eq:lepplanewaves}, as well as from the assumption of the exact energy conservation in Eq.~\eqref{eq:Smatrencons}. As we mentioned above, the plane waves and the energy conservation approximations are desirable but not crucial for our analysis. Nevertheless, the decoherence effects have not been studied in our work since we have adopted these approximations to advance in analytical calculations in Sec.~\ref{sec:TRANSPROB}.


We also make an interesting observation about neutrino oscillations
in external fields. The neutrino interaction with background matter
or with a magnetic field can be rather trivially incorporated to the
neutrino vacuum oscillations description. Indeed, one just adds a
linear term with a certain external field to the vacuum effective
Hamiltonian for flavor neutrinos. However, from the point of view
of QFT, the effect of external fields on neutrino oscillations is
an essentially nonperturbative phenomenon. The correct form of the
transition probability is obtained only if all terms in the infinite series in Eq.~(\ref{eq:Dysser}) are summed up.

\section*{Acknowledgments}

I am thankful to M.~Deka for the help with the Feynman diagram drawing.

\appendix

\section{Propagators of Weyl neutrinos in vacuum}

In this Appendix, we study the propagators of Majorana neutrinos in
vacuum. Using the Weyl form of the neutrino wavefunctions, we represent
the propagator for left-handed neutrinos and derive the propagator
for right-handed antineutrinos.

The propagator of the field $\eta_{a}$, which corresponds to left-handed
neutrinos, was obtained in Ref.~\cite{FukYan03} (see also Ref.~\cite{Dvo25},
where this propagator was derived in the presence of background matter).
Using this result, we represent the propagator in question in the
form,
\begin{equation}\label{eq:Seta}
  S_{a}^{(\eta)}(x)=\mathrm{i}\int\frac{d^{4}p}{(2\pi)^{4}}e^{-\mathrm{i}px}
  \frac{1}{2}
  \left\{
    \left[
      1-\frac{(\bm{\sigma}\mathbf{p})}{E_{a}}
    \right]
    \frac{1}{p_{0}-E_{a}+\mathrm{i}0}+
    \left[
      1+\frac{(\bm{\sigma}\mathbf{p})}{E_{a}}
    \right]
    \frac{1}{p_{0}+E_{a}-\mathrm{i}0}
  \right\},
\end{equation}
where $E_{a}=\sqrt{\mathbf{p}^{2}+m_{a}^{2}}$ and $\mathrm{i}0$
stays for a small imaginary term. We can decompose Eq.~\eqref{eq:Seta} as $S_{a}^{(\eta)}=S_{a\mathrm{L}}+\tilde{S}_{a\mathrm{R}}$,
where
\begin{equation}\label{eq:SL}
  S_{a\mathrm{L}}(p)=\frac{\mathrm{i}}{2}
  \left[
    1-\frac{(\bm{\sigma}\mathbf{p})}{E_{a}}
  \right]
  \frac{1}{p_{0}-E_{a}+\mathrm{i}0},
\end{equation}
is the propagator of left-handed neutrinos which is used in Sec.~\ref{sec:PROPB}.

The the wavefunction $\xi_{a}$, corresponding to a right handed antineutrino,
obeys the equation,
\begin{equation}\label{eq:xieq}
  \mathrm{i}\dot{\xi}_{a}-(\bm{\sigma}\mathbf{p})\xi_{a}-\mathrm{i}m_{a}\sigma_{2}\xi_{a}^{*}=0,
\end{equation}
which can be obtained from the Dirac equation for a Majorana neutrino
if we take that the bispinor has the form, $\psi_{a}^{\mathrm{T}}=(\xi_{a},-\mathrm{i}\sigma_{2}\xi_{a}^{*})$.
Such a bispinor satisfies the Majorana condition, $(\psi_{a})^{c}=\psi_{a}$.
The general solution of Eq.~(\ref{eq:xieq}) has the form,
\begin{align}\label{eq:gensolxi}
  \xi_{a}(\mathbf{x},t) = & \int\frac{\mathrm{d}^{3}p}{(2\pi)^{3/2}}\sqrt{\frac{E_{a}+p}{E_{a}}}
  \Big[
    \left(
      a_{a+}w_{+}+A_{a-}a_{a-}w_{-}
    \right)
    e^{-\mathrm{i}E_{a}t+\mathrm{i}\mathbf{px}}
    \notag
    \\
    & +
    \left(
      a_{a-}^{\dagger}w_{+}+A_{a+}a_{a+}^{\dagger}w_{-}
    \right)e^{\mathrm{i}E_{a}t-\mathrm{i}\mathbf{px}}
  \Big],
\end{align}
where
\begin{equation}
  A_{a\pm}=\pm\frac{m_{a}}{E_{a}+p},
\end{equation}
and
\begin{equation}\label{eq:helamp}
  w_{+}=
  \left(
    \begin{array}{c}
      e^{-\mathrm{i}\varphi/2}\cos\vartheta/2
      \\
      e^{\mathrm{i}\varphi/2}\sin\vartheta/2
    \end{array}
  \right),
  \quad
  w_{-}=
  \left(
    \begin{array}{c}
      -e^{-\mathrm{i}\varphi/2}\sin\vartheta/2
      \\
      e^{\mathrm{i}\varphi/2}\cos\vartheta/2
    \end{array}
  \right),
\end{equation}
are the helicity amplitudes, which satisfy the relation, $w_{\pm}\otimes w_{\pm}^{\dagger}=\frac{1}{2}\left[1\pm\bm{\sigma}\hat{p}\right]$.
In Eq.~(\ref{eq:helamp}), the angles $\vartheta$ and $\varphi$
fix the direction of the neutrino momentum $\mathbf{p}$. The creation
and annihilation operators, $a_{a\pm}^{\dagger}(\mathbf{p})$ and
$a_{a\pm}(\mathbf{p})$, obey the anticommurators,
\begin{equation}\label{eq:anticomm}
  \left\{
    a_{a\pm}(\mathbf{p}),a_{b\pm}^{\dagger}(\mathbf{q})
  \right\} _{+}
  =\delta_{ab}\delta(\mathbf{p}-\mathbf{q}),
\end{equation}
with the rest of the anticommutators being equal to zero.

Then, we define the propagator of $\xi_{a}$ as
\begin{equation}\label{eq:propxidef}
  S_{a}^{(\xi)}(x-y)=\theta(x_{0}-y_{0})
  \left\langle
    0
  \right|
  \xi(x)\xi^{\dagger}(y)
  \left|
    0
  \right\rangle
  -\theta(y_{0}-x_{0})
  \left\langle
    0
  \right|
  \xi^{*}(y)\xi^{\mathrm{T}}(x)
  \left|
    0
  \right\rangle.
\end{equation}
Using Eqs.~(\ref{eq:gensolxi})-(\ref{eq:anticomm}) and making the calculations
analogous to those in Refs.~\cite{FukYan03,Dvo25}, we cast Eq.~(\ref{eq:propxidef})
to the form,
\begin{equation}\label{eq:Sxi}
  S_{a}^{(\xi)}(x)=\mathrm{i}\int\frac{d^{4}p}{(2\pi)^{4}}e^{-\mathrm{i}px}\frac{1}{2}
  \left\{
    \left[
      1+\frac{(\bm{\sigma}\mathbf{p})}{E_{a}}
    \right]
    \frac{1}{p_{0}-E_{a}+\mathrm{i}0}+
    \left[
      1-\frac{(\bm{\sigma}\mathbf{p})}{E_{a}}
    \right]
    \frac{1}{p_{0}+E_{a}-\mathrm{i}0}
  \right\}.
\end{equation}
Similarly to $S_{a}^{(\eta)}$ in Eq.~(\ref{eq:Seta}), we decompose
$S_{a}^{(\xi)}$ in Eq.~(\ref{eq:Sxi}) as $S_{a}^{(\xi)}=S_{a\mathrm{R}}+\tilde{S}_{a\mathrm{L}}$,
where
\begin{equation}\label{eq:SR}
  S_{a\mathrm{R}}(p)=\frac{\mathrm{i}}{2}
  \left[
    1+\frac{(\bm{\sigma}\mathbf{p})}{E_{a}}
  \right]
  \frac{1}{p_{0}-E_{a}+\mathrm{i}0},
\end{equation}
is the propagator of right-handed antineutrinos which is used in Sec.~\ref{sec:PROPB}.


\begin{thebibliography}{50}

\bibitem{Wol78}
  L.~Wolfenstein,
  Neutrino Oscillations in Matter,
  Phys. Rev. D \textbf{17}, 2369--2374 (1978).

\bibitem{MikSmi85}
  S.~P.~Mikheyev and A.~Yu.~Smirnov,
  Resonance Amplification of Oscillations in Matter and Spectroscopy of Solar Neutrinos,
  Sov. J. Nucl. Phys. \textbf{42}, 913--917 (1985).

\bibitem{Xu23}
  X.-J.~Xu, Z.~Wang, and S.~Chen,
  Solar neutrino physics,
  Prog. Part. Nucl. Phys. \textbf{131}, 104043 (2023)
  [arXiv:2209.14832].

\bibitem{LeeSch77}
  B.~W.~Lee and R.~Shrock,
  Natural suppression of symmetry violation in gauge theories:
  Muon- and electron-lepton-number nonconservation,
  Phys. Rev. D \textbf{16}, 1444--1473 (1977).

\bibitem{Kin04}
  S.~F.~King,
  Neutrino Mass Models,
  Rept. Prog. Phys. \textbf{67}, 107--158 (2004)
  [hep-ph/0310204].

\bibitem{Giu24}
  C.~Giunti, K.~Kouzakov, Y.-F.~Li, and A.~Studenikin,
  Neutrino Electromagnetic Properties,
  Annu. Rev. Nucl. Part. Sci. \textbf{75} (2025),
  doi:~10.1146/annurev-nucl-102122-023242
  [arXiv:2411.03122].

\bibitem{LimMar88}
  C.-S.~Lim and W.~J.~Marciano,
  Resonant spin-flavor precession of solar and supernova neutrinos,
  Phys. Rev. D \textbf{37}, 1368--1373 (1988).

\bibitem{VolVysOku86}
  M.~B.~Voloshin, M.~I.~Vysotski\u\i, and L.~B.~Okun',
  Neutrino electrodynamics and possible consequences for solar neutrinos,
  Sov. Phys. JETP \textbf{64}, 446--452 (1986).

    
\bibitem{Akh88b}
  E.~Kh.~Akhmedov,
  Resonant Amplification of Neutrino Spin Rotation in Matter and the Solar Neutrino Problem,
  Phys. Lett. B \textbf{213}, 64--68 (1988).  

\bibitem{AkhPul03}
  E.~Kh.~Akhmedov and J.~Pulido,
  Solar neutrino oscillations and bounds on neutrino magnetic moment and solar magnetic field,
  Phys. Lett. B \textbf{553}, 7--17 (2003)
  [hep-ph/0209192].

\bibitem{SchVal81}
  J.~Schechter and J.~W.~F.~Valle,
  Majorana neutrinos and magnetic fields,
  Phys. Rev. D \textbf{24}, 1883--1889 (1981)
  [Erratum: ibid. \textbf{25}, 283 (1982)].

\bibitem{NauNau20}
  D.~V.~Naumov and V.~A.~Naumov,
  Quantum Field Theory of Neutrino Oscillations,
  Phys. Part. Nucl. \textbf{51}, 1--106 (2020).

\bibitem{Beu03}
  M.~Beuthe,
  Oscillations of neutrinos and mesons in quantum field theory,
  Phys. Rept. \textbf{375}, 105--218 (2003)
  [hep-ph/0109119].

\bibitem{Kob82}
  I.~Yu.~Kobzarev, B.~V.~Martem'yanov, L.~B.~Okun', and M.~G.~Shchepkin,
  Sum rules for neutrino oscillations,
  Sov. J. Nucl. Phys. \textbf{35}, 708--712 (1982).

\bibitem{GiuKimLee93}
  C.~Giunti, C.~W.~Kim, and J.~A.~Lee,
  On the treatment of neutrino oscillations without resort to weak eigenstates,
  Phys. Rev. D \textbf{48}, 4310--4317 (1993)
  [hep-ph/9305276].

\bibitem{GriSto96}
  W.~Grimus and P.~Stockinger,
  Real oscillations of virtual neutrinos,
  Phys. Rev. D \textbf{54}, 3414--3419 (1996)
  [hep-ph/9603430].

\bibitem{CarChu99}
  C.~Y.~Cardall and D.~J.~H.~Chung,
  MSW effect in quantum field theory,
  Phys. Rev. D \textbf{60}, 073012 (1999)
  [hep-ph/9904291].

\bibitem{AkhWil13}
  E.~Kh.~Akhmedov and A.~Wilhelm,
  Quantum field theoretic approach to neutrino oscillations in matter,
  J. High Energy Phys. \textbf{01}, 165 (2013)
  [arXiv:1205.6231].

\bibitem{EgoVol22}
  V.~Egorov and I.~Volobuev,
  Quantum field-theoretical description of neutrino oscillations in magnetic field,
  J. Exp. Theor. Phys. \textbf{135}, 197--208 (2022)
  [arxiv:2107.11570].

\bibitem{Dvo25}
  M.~Dvornikov,
  Quantum field theory treatment of neutrino flavor oscillations in matter,
  Phys. Rev. D \textbf{111}, 056009 (2025)
  [arxiv:2411.19120].

\bibitem{Dvo11}
  M.~Dvornikov,
  Field theory description of neutrino oscillations,
  in \textit{Neutrinos: Properties, Sources and Detection}, ed. by J.~P.~Greene
  (Nova Science Publishers, New York, 2011), pp.~23--90
  [arxiv:1011.4300].

\bibitem{Pas00}
  S.~Pastor, J.~Segura, V.~B.~Semikoz, and J.~W.~F.~Valle,
  Nucl. Phys. B \textbf{566}, 92--102 (2000)
  [hep-ph/9905405].

\bibitem{Dvo12}
  M.~Dvornikov,
  Evolution of a dense neutrino gas in matter and electromagnetic field,
  Nucl. Phys. B \textbf{855}, 760--773 (2012)
  [arxiv:1108.5043].

\bibitem{SchVal80}
  J.~Schechter and J.~W.~F.~Valle,
  Neutrino-oscillation thought experiment,
  Phys. Rev. D \textbf{23}, 1666--1668 (1980).

\bibitem{EgoLobStu00}
  A.~M.~Egorov, A.~E.~Lobanov, and A.~I.~Studenikin,
  Neutrino Oscillations in Electromagnetic Fields,
  Phys. Lett. B \textbf{491}, 137--142 (2000)
  [hep-ph/9910476].

\bibitem{BerLifPit82}
  V.~B.~Berestetski\u\i, E.~M.~Lifshitz, and L.~P.~Pitaevski\u\i,
  \textit{Quantum Electrodynamics}
  (Pergamon, Oxford, 1982), 2nd ed. 

\bibitem{GiuStu15}
  C.~Giunti and A.~Studenikin,
  Neutrino electromagnetic interactions: A window to new physics,
  Rev. Mod. Phys. \textbf{87}, 531 (2015)
  [arXiv:1403.6344]

\bibitem{BlaVit96}
  M.~Blasone and G.~Vitiello,
  Quantum field theory of fermion mixing,
  Ann. Phys. (N.Y.) \textbf{244}, 283--311 (1996);
  Erratum: ibid. \textbf{249}, 363--364 (1996)
  [hep-ph/9501263].

\bibitem{Bla23}
  M.~Blasone, F.~Giacosa, L.~Smaldone, and G.~Torrieri,
  Neutrino oscillations in the interaction picture,
  Eur. Phys. J. C \textbf{83}, 736 (2023)
  [arXiv:2305.07107].

\bibitem{BlaSma25}
  M.~Blasone and L.~Smaldone,
  Perturbative and nonperturbative aspects of neutrino oscillations in quantum field theory,
  Int. J. Mod. Phys. A \textbf{40}, 2530017 (1996)
  [arXiv:2508.18917].  
  
\bibitem{Lob19}
  A.~E.~Lobanov,
  Particle quantum states with indefinite mass and neutrino oscillations,
  Ann. Phys. (N.Y.) \textbf{403}, 82--105 (2019)
  [arXiv:1507.01256].

\bibitem{Ans23}
  D.~Anselmi,
  Quantum Field Theory of Physical and Purely Virtual Particles in a Finite Interval of Time on a Compact Space Manifold:
  Diagrams, Amplitudes and Unitarity,
  J. High Energy Phys. \textbf{07}, 209 (2023)
  [arXiv:2304.07642].

\bibitem{Tur23}
  A.~Tureanu,
  Neutrino oscillations by a manifestly coherent mechanism and massless vs. massive neutrinos,
  Phys. Lett. B \textbf{843}, 137996 (2023)
  [arXiv:2304.13491].

\bibitem{AkhKop10}
  E.~Kh.~Akhmedov and J.~Kopp,
  Neutrino oscillations: Quantum mechanics vs. quantum field theory,
  J. High Energy Phys. \textbf{04}, 008 (2010)
  [arXiv:1001.4815].

\bibitem{GriStoMoh99}
  W.~Grimus, P.~Stockinger, and S.~Mohanty,
  The field theoretical approach to coherence in neutrino oscillations,
  Phys. Rev. D \textbf{59}, 013011 (1999)
  [hep-ph/9807442].


\bibitem{FukYan03}
  M.~Fukugita and T.~Yanagida,
  \textit{Physics of Neutrinos and Applications to Astrophysics}
  (Springer, Berlin, 2003), pp.~292--296.

\end{thebibliography}
\end{document}